\documentclass[preprint,aps,prb]{revtex4}
\usepackage{graphicx,amsmath,bbm} 
\date{\today}

\begin{document}

\title{Dynamic treatment of vibrational energy relaxation \\
in a heterogeneous and fluctuating environment}

\author{Hiroshi Fujisaki}\email{fujisaki@theochem.uni-frankfurt.de} 
\author{Gerhard Stock}\email{stock@theochem.uni-frankfurt.de} 
\affiliation{Institute of Physical and Theoretical Chemistry,
J.\ W.\ Goethe University, Max-von-Laue-Str.\ 7,
60438 Frankfurt am Main, Germany}

\begin{abstract}
A computational approach to describe the energy relaxation of a
high-frequency vibrational mode in a fluctuating heterogeneous
environment is outlined. Extending previous work [H.~Fujisaki,
Y.~Zhang, and J.E.~Straub, J.~Chem.~Phys.~{\bf 124}, 144910 (2006)],
second-order time-dependent perturbation theory is employed which
includes the fluctuations of the parameters in the Hamiltonian within
the vibrational adiabatic approximation. This means that the
time-dependent vibrational frequencies along an MD trajectory are
obtained via a partial geometry optimization of the solute with fixed
solvent and a subsequent normal mode calculation.  Adopting the amide
I mode of N-methylacetamide in heavy water as a test problem, it is
shown that the inclusion of dynamic fluctuations may significantly
change the vibrational energy relaxation. In particular, it is found
that relaxation occurs in two phases, because for short times
($\lesssim$ 200 fs) the spectral density appears continuous due to the
frequency-time uncertainty relation, while at longer times the
discrete nature of the bath becomes apparent. Considering the
excellent agreement between theory and experiment, it is speculated if
this behavior can explain the experimentally obtained biphasic
relaxation the amide I mode of N-methylacetamide.
\end{abstract}


\maketitle

%
%
\section{Introduction}

Vibrational energy relaxation (VER) is ubiquitous in chemistry and
physics. When a chemical reaction or conformational change occurs,
some vibrational modes may be excited or de-excited, resulting in
nonequilibrium phenomena of VER (for example, see
Ref.~\onlinecite{Agarwal} for the possible role of VER in enzyme
reactions). It is usually assumed that VER is much faster than
chemical reactions or conformational changes,\cite{Steinfeldbook} but
this is not always the case.\cite{LW97,GW04} Recent progress in
time-resolved spectroscopy\cite{MillerReview,
MK97,HLH98,ZAH01,Tokmakoff06, FayerReview,DRID00,NNT05,Hammreview} as
well as numerous theoretical formulations \cite{Oxtoby,KTF94,
Straub,WWH92,Henry,Straub2,NS03,Moritsugu2000,Nagaoka,Okazaki,Geva,FBS05,
Leitner05,DJBK07,Skinner,Hynes,Sibert,Marx04,FZS06,FS07,FYHS07,FYHSS08}
have been trying to unravel the molecular origin of VER.

To define the problem, we consider the case of the photoinduced $n=
1\!\rightarrow \!0$ VER of a high-frequency ($\approx$ 1000$-$2000
cm$^{-1}$) vibrational mode in a polyatomic molecule. We partition the
total Hamiltonian $H$ as
\begin{equation} \label{Htot}
H = H_{ S} +H_{ B} +H_{ SB},
\end{equation}
where the system Hamiltonian $H_{ S}$ describes the high-frequency
mode $\omega_S$, the bath Hamiltonian $H_{ B}$ comprises all
remaining vibrational modes $\omega_\alpha$ of the molecule, and
$H_{ SB}$ accounts for the coupling between the system and bath. 
The dynamics of the total system is described by the Liouville
equation 
\begin{equation} \label{LvN}
{\rm i} \hbar \frac{\partial \rho(t)}{\partial t} =
[H,\rho(t)]
\end{equation}
with $\rho(t)$ being the density operator for the system and bath.
Employing fully quantum-mechanical formulations, various
quantum-classical approaches or quasiclassical methods, a number of
groups have described VER by calculating the time evolution of the
total system.
While this strategy is formally exact, it is also quite
time-consuming and therefore limited to small molecules or model
systems. Thus, various reduced density-matrix formulations for the
system part in Eq.~(\ref{LvN}) have been pursued in the fields of
quantum optics, condensed matter physics, and physical
chemistry.\cite{Weiss,Breuer,Nitzan}

Considering VER, it is often the case that the coupling $V = H_{ SB} -
\langle H_{ SB} \rangle_B$ (where $\langle ... \rangle_B$ denotes the
bath average) is small enough to be described by low-order
time-dependent perturbation theory. Assuming, for simplicity, that the
VER is dominated by cubic coupling $C_{S \alpha \beta}$ between the
system mode $q_S$ and two bath modes $q_{\alpha}, q_{\beta}$ (i.e.,
$V \propto \sum_{\alpha \beta} C_{S \alpha \beta} q_S q_{\alpha}
q_{\beta}$), we obtain Fermi's Golden Rule\cite{Oxtoby,KTF94}
\begin{equation} \label{GR}
k_{1\rightarrow 0} \propto \sum_{\alpha,\beta} 
\left|\langle f|V|i \rangle \right|^2 
\delta (\omega_S -\omega_\alpha - \omega_\beta) 
\end{equation}
for the VER rate from the vibrationally excited state $|1 \rangle$ to
the ground state $|0 \rangle$, where $|i \rangle$ and $|f \rangle$
denote the initial and final state of the total system,
respectively. If we are only interested in the short-time decay of our
initially prepared state (rather than the time evolution of the
complete system as in Eq.\ (\ref{LvN})), Fermi's Golden Rule provides
a convenient and correct description of VER. However, in many cases
(e.g., in condensed phase VER) the straightforward implementation of
the Golden Rule is hampered by the fact that the true form of the bath
Hamiltonian $H_{ B}$ and the coupling Hamiltonian $H_{ SB}$ is
not known.  As a remedy, either idealized models (e.g., a harmonic
bath with bilinear couplings to the system mode) or classical
molecular dynamics (MD) simulations have been invoked, which allow us
to directly calculate the spectral density of the bath and its
(inverse) Fourier transform, the bath correlation function $\langle V(t) V(0)
\rangle$. Rewriting Eq.\ (\ref{GR}) in its time-dependent form, the
Golden Rule can be directly expressed in terms of the bath correlation
function
\begin{equation} \label{GR2}
k_{1\rightarrow 0} \propto 
\int^{\infty}_{-\infty} dt \,e^{{\rm i}\Delta \omega_{fi}t}\,
\langle V(t) V(0) \rangle , 
\end{equation}
where $\Delta \omega_{fi}$ accounts for the frequency difference of
the initial and final state. In the case of high-frequency modes, a
classical bath correlation function (sampled at $k_B T$) may be only a
poor approximation to the quantum correlation function (dominated by
zero point energy motion). Hence it is a well-established approach to
augment Eq.\ (\ref{GR2}) with quantum correction
factors.\cite{Skinner,Hynes,Sibert,Marx04} Another extension of 
Eq.~(\ref{GR2}) was given by Bakker, who explicitly included the
time-dependent fluctuations of the bath frequencies \cite{Bakker04}.

Recently, Fujisaki {\it et al.}\cite{FZS06,FS07} proposed an
alternative approach to include in a realistic manner the effects of
an inhomogeneous environment into the description of VER. The idea is
to first select numerous snapshots of the solute molecule and its
surrounding solvent shell from an equilibrium MD trajectory. For each
structure, instantaneous normal mode analysis of the solvated system
is performed, which gives the instantaneous vibrational frequencies
$\omega_S$ and $\omega_\alpha$ of this conformation as well as the
instantaneous vibrational couplings. Subsequently, a time-dependent
perturbative calculation (which can reduce to the Golden Rule) is
performed for each snapshot, and the overall rate is obtained by a
direct average over all conformation-dependent rates. The approach
rests on the assumptions (i) that instantaneous normal mode
calculations provide a correct representation of the time-dependent
vibrational frequencies and (ii) that these frequencies are constant
on the time scale of the VER process, thus rendering a direct
inhomogeneous averaging appropriate. A similar treatment using the
path-integral method has been proposed by Shiga, Okazaki and their
coworkers.\cite{Okazaki}

The goal of this work is to go beyond these assumptions by explicitly
considering the time-dependence of instantaneous vibrational
frequencies and couplings during the VER process. To this end, we
include the time-dependent driving of the environment into the
perturbative formulation of Refs.~\onlinecite{FZS06,FS07}. Adopting
N-methylacetamide (NMA) in heavy water as a simple peptide
model to study the VER of the amide I vibration, we discuss the effects
of the various treatments of the fluctuations. Furthermore, we show
that, within the adiabatic approximation underlying the approach, the
time-dependent vibrational frequencies need to be obtained from
geometry-optimized (rather than true instantaneous) molecular
structures. The calculations nicely reproduce the subpicosecond VER of
the amide I mode in NMA measured in recent transient infrared experiments.
\cite{HLH98,ZAH01,Tokmakoff06}

\newpage
%
%
\section{Theory and methods} \label{sec:method}
\subsection{Perturbation theory: Previous formulation}

Before considering the time-dependent driving, it is helpful to first
briefly summarize the derivation of the perturbative formulation of
VER given by Fujisaki {\it et al}.\cite{FZS06,FS07}
The total Hamiltonian (\ref{Htot}) is partitioned as $H=H_0+V$ with
\begin{eqnarray}
H_0 &=& H_{S} + H_{B} + \langle H_{SB}  \rangle_{B}, \\
V   &=& H_{SB} - \langle H_{SB}  \rangle_{B},
\end{eqnarray}
where $\langle \ldots \rangle_{B}$ denotes the average over the bath
degrees of freedom. We have defined the interaction Hamiltonian $V$
such that $\langle V \rangle_B=0$, which is appropriate for the
perturbative treatment.\cite{Hanggi} Employing time-dependent
perturbation theory with respect to the coupling $V$, the system
density operator $\rho_{S}(t)= {\rm Tr_B} \rho(t)$ in second order can
be written as
\begin{equation} \label{rho1}
\rho_{S}^{(2)}(t) = \frac{1}{(i \hbar)^2} {\rm Tr_B} 
\int^{t}_0 dt_2 \int^{t_2}_0 dt_1 U_{S}(t)
\left[ V(t_2), \left[ V(t_1), \rho(0)\right] \right]
U_{S}^{\dagger}(t) .
\end{equation}
Here $V(t)=U_0^\dagger(t) V U_0(t)$ represent the coupling $V$ in the
interaction representation and we have introduced unperturbed 
propagators $U_k(t)=e^{-iH_k t/\hbar}$ with $k=0,{S,B}$.

To be specific, in the following we assume that the system and the
bath is described in the harmonic approximation
\begin{eqnarray}
H_{S} &=& \frac{p_S^2}{2} + \frac{\omega_S^2}{2} q_S^2, \\
H_{B} &=& \sum_{\alpha} \left( \frac{p_\alpha^2}{2} 
+ \frac{\omega_{\alpha}^2}{2} q_\alpha^2 \right),
\end{eqnarray}
where $p_i$ and $q_i$ ($i=S,\alpha$) are momenta and
positions of the $i$th vibrational mode with frequency $\omega_i$.
Furthermore, we assume that the VER is dominated by cubic coupling
between the system mode and two bath modes, resulting in
\begin{equation}
H_{SB} = - q_{S} F = q_{S}\sum_{\alpha,\beta} 
C_{S \alpha \beta} q_\alpha q_\beta ,
\label{eq:force}
\end{equation}
which noticeably differs from a bilinear coupling commonly
employed.\cite{Weiss,Breuer,Nitzan,IT06} We consider the case that at time $t=0$ the
total density operator factorizes in system and bath operator, i.e.,
$\rho(0) = \rho_{S}(0) \rho_{B}(0)$, and that initially the system is
in its first excited state, i.e., $\rho_{S}(0) =|1 \rangle \langle
1|$.  
Combining Eqs.\ (\ref{rho1}) - (\ref{eq:force}), the population
probability of the amide I vibrational ground state can be
written as
\begin{eqnarray}
\rho^{(2)}_{00}(t) 
&\equiv& \langle 0| \rho_S^{(2)}(t) |0 \rangle 
\nonumber
\\
&=& \frac{2}{\hbar^2} |\langle 0| q_S |1 \rangle |^2
\int^{t}_0 dt_2 \int^{t_2}_0 dt_1 
\left[
\langle \delta F(t_2-t_1) \delta F(0) \rangle_B \;
e^{i \omega_S(t_2-t_1)}
+ c.c. \right]
\label{rho2}
\end{eqnarray}
with $\delta F(t) = F(t) - \langle F \rangle_B$ and
$F(t)=U_B^\dagger(t) F U_B(t)$.  Assuming, for simplicity, that only
bath modes with $\hbar \omega_{\alpha} > k_B T \simeq 200$ cm$^{-1}$
are of importance in the VER of amide I vibrations, the bath
correlation function can be written as \cite{FBS05}
\begin{equation}
\langle \delta F(t_2) \delta F(t_1) \rangle_B = 
\frac{\hbar^2}{2} \sum_{\alpha,{\beta}} 
\frac{C_{S \alpha \beta}^2}{\omega_{\alpha} \omega_{\beta}}
e^{-i (\omega_{\alpha}+\omega_{\beta})(t_2-t_1)}\; .
\end{equation}
Insertion into Eq.\ (\ref{rho2}) yields the final result for the
time-dependent ground state population
\begin{equation} \label{eq:VERexpo2}
\rho^{(2)}_{00}(t)
= \frac{\hbar}{2 {\omega}_S} \sum_{\alpha,\beta} 
\frac{C_{S \alpha \beta}^2}{\omega_{\alpha} \omega_{\beta}}\;
\frac{[1-\cos({\omega}_S-\omega_{\alpha}-\omega_{\beta})t]}
{({\omega}_S-\omega_{\alpha}-\omega_{\beta})^2}\; .
\end{equation}
To recover the corresponding Golden Rule expression (\ref{GR}), the
long-time limit of $d \rho_{00}/dt$ is taken.\cite{Oxtoby} 
Following Ref.\
\onlinecite{FZS06}, however, here we define the VER probability as
\begin{equation} \label{eq:VERexpo7}
P(t) \equiv  e^{-\rho^{(2)}_{00}(t)} .
\end{equation}
At short times $t\lesssim$ 300 fs, $P(t) \approx 1\! -\!
\rho^{(2)}_{00}(t)$.  At longer times, Eq.\ (\ref{eq:VERexpo7}) may be
advantageous, since it avoids problems associated with the violation
of positivity of the density matrix (see below).

Finally, we account for the inhomogeneity of the environment by
evaluating the VER probability (\ref{eq:VERexpo7}) for a number of MD
snapshots $\{ r=1,\ldots,N \}$. Each snapshot provides instantaneous
vibrational frequencies $\omega_{S}^{(r)}(t)$, $\omega_{\alpha}^{(r)}(t)$ and
couplings $C_{S \alpha \beta}^{(r)}(t)$ of this conformation. The mean
relaxation probability $\langle P(t) \rangle_S $ is then obtained by a
statistical average over all conformation-dependent probabilities
$P_r(t)$
\begin{equation} \label{eq:VERexpo5}
\langle P(t) \rangle_S  = \frac{1}{N} \sum_{r=1}^N P_r(t) \,.
\end{equation}
Alternatively, we may approximate the mean relaxation probability by a
second-order cumulant expansion, yielding
\begin{eqnarray} \label{eq:VERexpo6}
\langle P(t) \rangle_S  &=& \left\langle 1 - \left[\rho^{(2)}_{00}(t) +
\rho^{(4)}_{00}(t) + \ldots \right] \right\rangle_S \nonumber \\
&\approx& \exp \left\{ - \left\langle \rho^{(2)}_{00}(t) \right\rangle_S
\right\} \, .
\end{eqnarray}

\begin{figure}[htbp]
\hfill
\begin{center}
\includegraphics[scale=1.0]{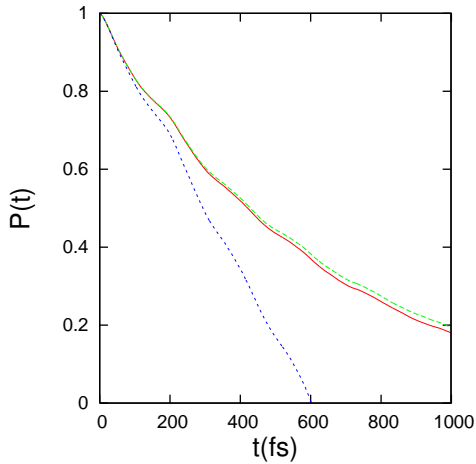}
\end{center}
\caption{\baselineskip5mm
Mean vibrational energy relaxation probability $\langle P(t)
  \rangle_S$ as obtained for NMA in D$_2$O. Compared is the outcome of
  direct averaging [Eqs.\ (\ref{eq:VERexpo5}) and (\ref{eq:VERexpo7}),
  red line], cumulant approximation [Eq.\ (\ref{eq:VERexpo6}), green
  line] and the naive ansatz $\langle P(t) \rangle_S = 1 -
  \left\langle \rho^{(2)}_{00}(t) \right\rangle_S$ (blue line).}
\label{compPt}
\end{figure}

Figure \ref{compPt} compares the outcome of the various possible
definitions of the mean VER probability.  [Results are obtained from
partial energy minimization and dynamical averaging, see below.]  We
find that the direct averaging [Eqs.\ (\ref{eq:VERexpo5}) and
(\ref{eq:VERexpo7})] and the cumulant approximation [Eq.\
(\ref{eq:VERexpo6})] give virtually the same results, while the naive
ansatz $\langle P(t) \rangle_S = 1 - \left\langle \rho^{(2)}_{00}(t)
\right\rangle_S$ deviates after $\approx$ 300 fs and eventually
becomes negative. The latter is associated with the fact that the
eigenvalues of a reduced density matrix can be
negative.\cite{positivity} Note that Figure \ref{compPt} also
motivates our ansatz (\ref{eq:VERexpo7}) for the VER probability of
individual trajectories. As the cumulant approximation appears to
provide the clearest derivation of the mean VER probability, in the
following we will use this definition for $\langle P(t) \rangle_S$.

%
%
\subsection{New formulation: Time-dependent driving} \label{sec:tdmethod}

In the perturbation theory described above, the vibrational
frequencies and coupling elements are assumed to be constant {\em
  during} the VER process.\cite{FZS06,FS07} This is an assumption
which may break down if the time scale of the VER is similar to that
of the vibrational parameters.  In this work, we extend this
formulation by explicitly considering the time dependence of
instantaneous vibrational frequencies $\omega_{S} = \omega_{S}(t)$ and
$\omega_{\alpha} = \omega_{\alpha}(t)$ and coupling elements $C_{S
  \alpha \beta} = C_{S \alpha \beta}(t)$ in the derivation of the
population probability. As a consequence, the Hamiltonian of system
and bath become time-dependent, $H_S=H_S(t)$ and $H_B=H_B(t)$, thus
yielding the corresponding propagators
\begin{eqnarray}
U_S(t)  &=& e^{-(i/\hbar) \int_0^t H_S(\tau) d \tau } \,,
\\
U_B(t)  &=& e^{-(i/\hbar) \int_0^t H_B(\tau) d \tau } \,.
\end{eqnarray}
Second-order time-dependent perturbation theory then leads to an
expression that is formally quite similar to Eq.~(\ref{rho2}), 
except that the time dependence of all vibrational parameters
is considered.
The resulting bath correlation function reads 
\begin{eqnarray} \label{cf}
\langle \delta F(t_2) \delta F(t_1) \rangle_B &=& 
\frac{\hbar^2}{2} \sum_{\alpha,{\beta}} 
\frac{C_{S \alpha \beta}(t_2)C_{S \alpha \beta}(t_1)}
{\sqrt{\omega_{\alpha}(t_2) \omega_{\beta}(t_1)
\omega_{\alpha}(t_2) \omega_{\beta}(t_1)}} 
\nonumber \\
&& \times
\exp \left \{
-i \int_{t_1}^{t_2} [\omega_{\alpha}(\tau)+\omega_{\beta}(\tau)]d \tau
\right \} ,
\end{eqnarray}
and we obtain for the time-dependent ground-state population probability
\begin{eqnarray} \label{eq:VERexpo3}
\rho^{(2)}_{00}(t) &=& \frac{\hbar}{2} \sum_{\alpha,\beta}
 \int^{t}_0 dt_2 \int^{t_2}_0 dt_1 
\frac{C_{S \alpha \beta}(t_2)C_{S \alpha \beta}(t_1)}
{\sqrt{\omega_{S}(t_2) \omega_{S}(t_1)
\omega_{\alpha}(t_2) \omega_{\beta}(t_1)
\omega_{\alpha}(t_2) \omega_{\beta}(t_1)}} 
\nonumber \\
&& \times
\cos \int_{t_1}^{t_2} [\omega_S(\tau) -\omega_{\alpha}(\tau)
- \omega_{\beta}(\tau)] d \tau \; .
\end{eqnarray}
The mean VER probability in Eq.\
(\ref{eq:VERexpo5}) is again defined by averaging over all
conformation-dependent probabilities obtained from
various MD trajectories. Equation (\ref{eq:VERexpo3}) represents the
main theoretical result of this paper. We refer to this result 
as ``dynamic treatment'' of VER, because it takes into account the time-dependent
fluctuations of the environment. Assuming constant frequencies and
couplings, we of course recover Eq.~(\ref{eq:VERexpo2}), which can be
regarded as the inhomogeneous limit of Eq.~(\ref{eq:VERexpo3}).

%
%
\subsection{Calculation of time-dependent frequencies}

In the formulation of VER developed above, we need to calculate
instantaneous vibrational frequencies $\omega_i(t)$ and couplings
$C_{S \alpha \beta}(t)$ along an MD trajectory. As the VER of the
amide I mode also involves the vibrations of the first water shell of
the peptide, we include NMA and the 16 nearest water molecules in the
normal mode analysis (see Sec.~\ref{sec:compmethod} below; the
effect of the size of the included solvent shell is discussed in
Fig.~\ref{fig:comp}). There are several ways to calculate the
normal modes of this subsystem for an instantaneous molecular
conformation. As adopted in Refs.\ \onlinecite{FZS06,FS07}, for
example, one may simply perform an instantaneous normal mode
calculation at this molecular geometry. Alternatively, one may  first
perform a partial geometry optimization of {NMA} (i.e., keeping 
the 16 solvent water fixed), before the normal mode calculation is done.

\begin{figure}[htbp]
\hfill
\begin{center}
\includegraphics[scale=0.4]{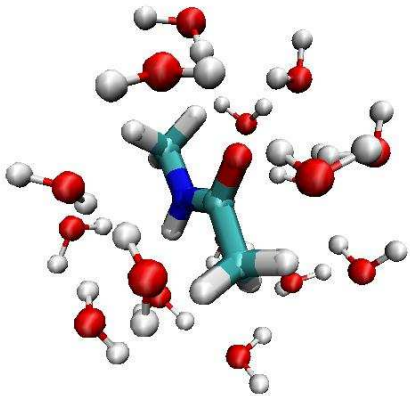}
\end{center}
\caption{\baselineskip5mm
MD snapshot of N-methylacetamide (NMA) and
its 16 nearest water molecules.}
\label{fig:nma}
\end{figure}

The latter procedure is sometimes referred to as ``quenched normal
modes'' method,\cite{OT90} and is based on the underlying adiabatic
approximation for the calculation of the (high-frequency) amide I
mode. To explain this, we consider the standard Born-Oppenheimer
approach to molecular electronic structures. Here the nuclear (i.e.,
slow) degrees of freedom coordinates are fixed, while the Schr\"odinger
equation for the electronic (i.e., fast) degrees of freedom is
solved. In a direct analogy, to calculate the structure of
high-frequency vibrational modes, we keep the slow solvent degrees of
freedom fixed and solve the problem of the fast degrees of
freedom. Within the harmonic approximation, the latter amounts to a
geometry optimization of the subsystem followed by a normal mode
calculation.

%
%
\subsection{Simulation details} \label{sec:compmethod}

All simulations were performed using the CHARMM simulation program
package.\cite{CHARMM} We employed the CHARMM22 all-atom force
field\cite{CHARMMFF} to model the solute NMA (H$_3$C-COND-CH$_3$) and the
TIP3P water model\cite{TIP3P} with doubled hydrogen masses to model the
solvent D$_2$O. We also performed simulations for fully deuterated NMA
(D$_3$C-COND-CD$_3$). The peptide was placed in a periodic cubic box
of (25.5 \AA)$^3$ containing 551 D$_2$O molecules. All bonds
containing hydrogen bonds were constrained by SHAKE algorithm
\cite{SHAKE} with a relative geometric tolerance of 10$^{-9}$.  
We used a 10 \AA \, cutoff with a switching function for the nonbonded
interaction calculations. After a standard equilibration protocol, we
ran a 100 ps NVT trajectory at 300 K, from which 100 statistically
independent configurations were sampled. For each initial condition, a
1 ps NVE run was performed for the VER calculations, using a time step
of 1 fs.

For the normal mode calculations, the Hessian matrix 
{with respect to the mass-weighted Cartesian coordinates $x_i$} 
\begin{equation}
K_{ij} =
\frac{1}{\sqrt{m_i m_j}} \frac{\partial^2 V} {\partial x_i \partial x_j} 
\end{equation}
of the system NMA/(D$_2$O)$_{16}$
was calculated by using the {\tt vib} command of CHARMM. 
We obtain in total 180 normal modes.
In the partial minimization scheme, the {\tt cons fix sele} command
was employed to constrain all atoms of water except NMA. The {\tt min}
command was used to minimize the energy of the subsystem NMA.  The
cubic couplings $C_{S \alpha \beta}$ with respect to the normal modes
$q_{i}$ were calculated from the Hessian matrix using numerical
differentiation\cite{FBS05}
\begin{eqnarray} \label{eq:CSab}
C_{S \alpha \beta} &=& - \frac{1}{2} 
\frac{\partial^3 V}{\partial q_S \partial q_\alpha \partial q_\beta} 
\nonumber \\
&\simeq& - \frac{1}{2} \sum_{ij}
U_{i \alpha}U_{j \beta} 
\frac{K_{ij}(\Delta q_S)-K_{ij}(-\Delta q_S)}{2 \Delta q_S} ,
\end{eqnarray}
where $\{U_{i \alpha}\}$ comprises the eigenvectors of the Hessian
matrix. To avoid problems with low-frequency bath modes whose
frequency may become zero (or even imaginary in the case of
instantaneous normal modes) in the denominator of Eqs.\
(\ref{eq:VERexpo2}) and (\ref{eq:VERexpo3}), we only include bath modes
with $\omega_{\alpha} \ge$ 100 cm$^{-1}$.

\newpage
%
%
\section{Computational results} \label{sec:results}
\subsection{Time-dependent frequencies}

\begin{figure}[htbp]
\hfill
\begin{center}
\begin{minipage}{.42\linewidth}
\includegraphics[scale=.7]{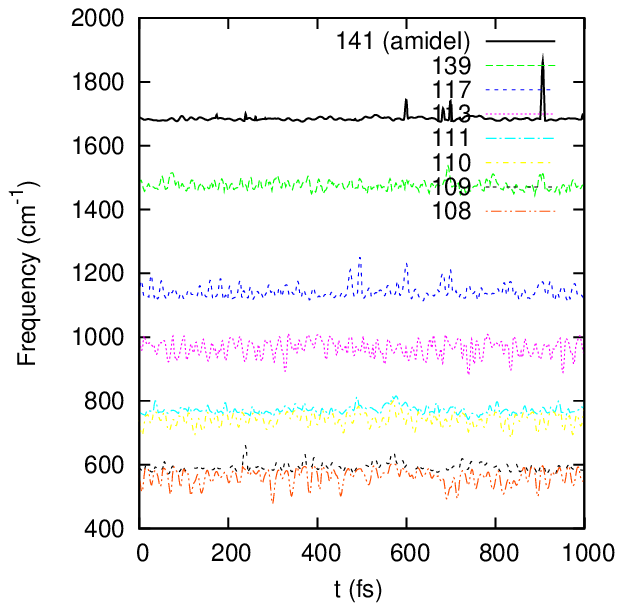}
\end{minipage}
\hspace{-2cm}
\begin{minipage}{.42\linewidth}
\includegraphics[scale=.7]{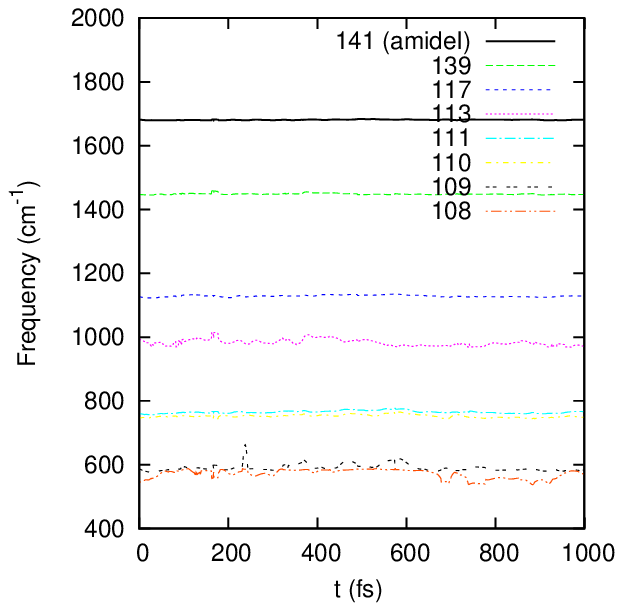}
\end{minipage}
\begin{minipage}{.42\linewidth}
\includegraphics[scale=.7]{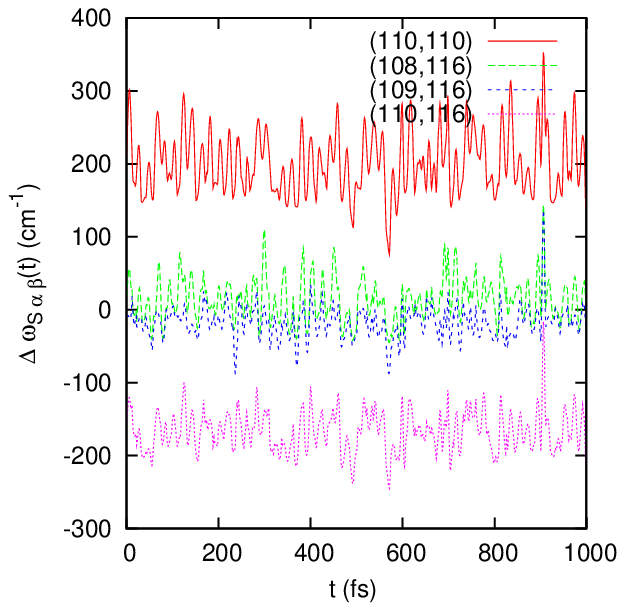}
\end{minipage}
\hspace{-2cm}
\begin{minipage}{.42\linewidth}
\includegraphics[scale=.7]{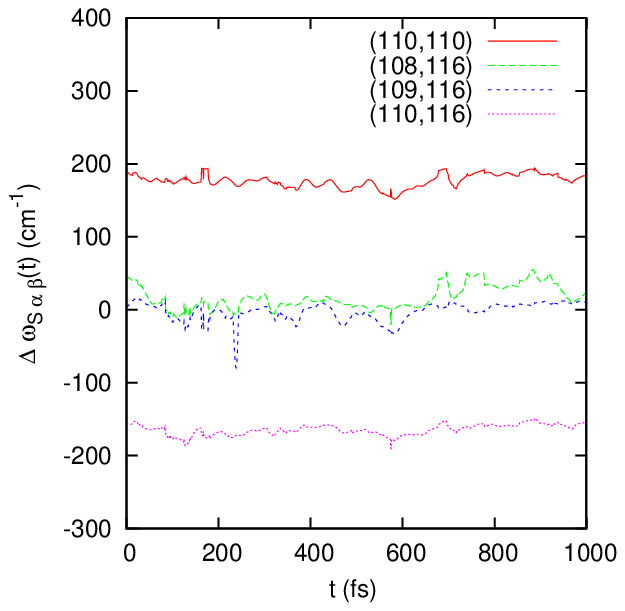}
\end{minipage}
\begin{minipage}{.42\linewidth}
\includegraphics[scale=.7]{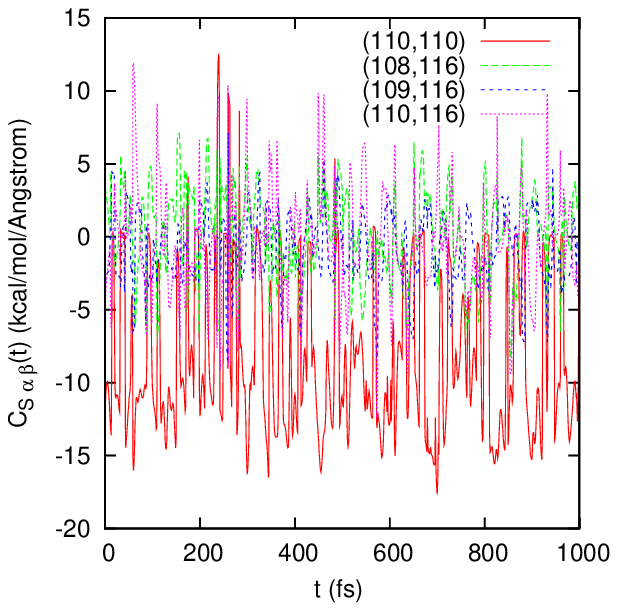}
\end{minipage}
\hspace{-2cm}
\begin{minipage}{.42\linewidth}
\includegraphics[scale=.7]{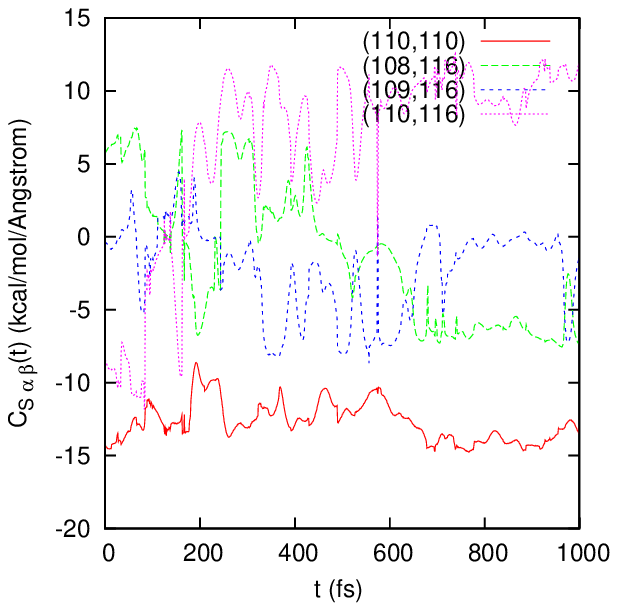}
\end{minipage}
\end{center}
\caption{\baselineskip5mm
Time evolution of the vibrational dynamics of
  NMA in D$_2$O, as obtained from instantaneous
  normal mode analysis with (right) and without (left) partial energy
  minimization. Shown are (upper panels) the normal mode frequencies
  $\omega_i$ of selected vibrational modes, (middle panels) the
  frequency mismatch $\Delta \omega_{S \alpha \beta}(t) = \omega_S(t)
  -\omega_{\alpha}(t) - \omega_{\beta}(t)$ for several resonant bath
  mode combinations, and (lower panels) the corresponding third-order
  anharmonic couplings $C_{S \alpha \beta}(t)$.}
\label{fig:amideI}
\end{figure}

Let us first consider the time-dependent vibrational frequencies along
a typical trajectory by performing instantaneous normal mode analysis
of NMA and its 16 nearest water molecules. Figure \ref{fig:amideI}
(top panels) shows the time evolution of the amide I mode (mode 141)
and several other modes that are of importance in the VER of the amide
I mode (see Table \ref{table:norm1} below). The vibrational
frequencies obtained from the normal mode calculations at the
instantaneous (not minimized) molecular structures (left panel) are
seen to undergo substantial fluctuations. The range of fluctuations
appears quite unrealistic, when compared to typical experimental
infrared line widths. The right panel of Fig.~\ref{fig:amideI} shows
results obtained from normal mode calculations for partially minimized
molecular structures. We see that the partial geometry optimization of
the subsystem NMA/(D$_2$O)$_{16}$ results in a reduction of the
frequency fluctuations to a physically reasonable range. This finding
indicates -- in accordance with the underlying adiabatic
approximation -- that normal mode calculations should be performed
after partial geometry optimization rather than at the instantaneous
molecular structure. The same conclusion was reached in a recent study
of the VER of isolated NMA on ab initio potential energy
surfaces.\cite{FYHSS08}

To assess the effect of the frequency fluctuations on the VER of the
amide I mode, we next consider the time dependence of the resonance
condition $\Delta \omega_{S \alpha \beta}(t) = \omega_S(t)
-\omega_{\alpha}(t) - \omega_{\beta}(t)$. As is evident from
Eqs.\ (\ref{eq:VERexpo2}) and (\ref{eq:VERexpo3}), the minima of
$|\Delta \omega_{S \alpha \beta}(t)|$ largely determine the efficiency
of the VER process. Adopting the most important mode combinations
$\{S, \alpha, \beta \}$ in the VER of the amide I mode of NMA (see
Table \ref{table:FRP1} below), Fig.\ \ref{fig:amideI} (middle panels)
compares the time evolution of $\Delta \omega_{S \alpha \beta}(t)$
obtained from normal mode calculations with (right) and without (left)
partial geometry optimization of the subsystem NMA. The considerable
differences found for the two results of $\Delta \omega_{S \alpha
  \beta}(t)$ again underlines the importance of using the appropriate
(i.e., partially minimized) normal mode frequencies.

Also shown in Fig.\ \ref{fig:amideI} are the cubic couplings $C_{S
\alpha \beta}(t)$ associated with the VER process. From their
definition in Eq.\ (\ref{eq:CSab}), it is clear that this quantity, too,
depends on how the normal modes are calculated. As expected, we find
that the results for $C_{S \alpha \beta}(t)$ without minimization 
fluctuate much more compared to the results after the partial
minimization. We note that apart from the frequencies we also
need the normal mode eigenvectors to calculate $C_{S \alpha
\beta}(t)$. For simplicity, however, we have assumed that the character
of the modes does not change significantly during the VER process and
therefore neglected the explicit time dependence of the eigenvectors.
This assumption may break down, when it is applied to
strongly fluctuating instantaneous normal modes.

%
%
\subsection{Numerical evaluation of the dynamic VER formula}

\begin{figure}[htbp]
\hfill
\begin{center}
\begin{minipage}{.42\linewidth}
\includegraphics[scale=.7]{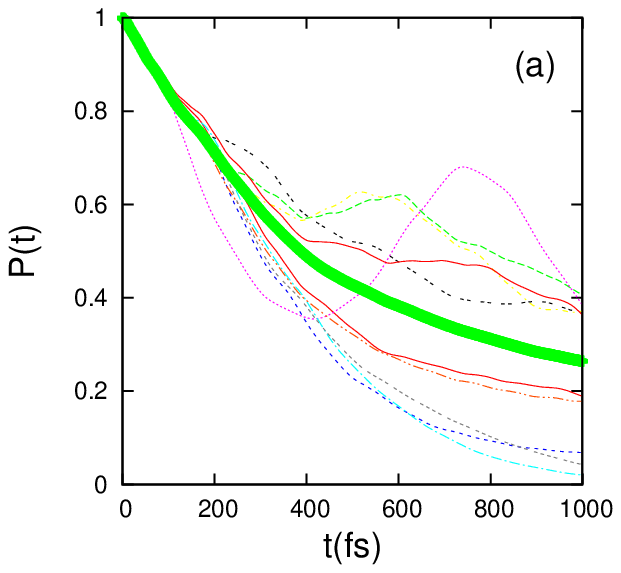}
\end{minipage}
\hspace{-2cm}
\begin{minipage}{.42\linewidth}
\includegraphics[scale=.7]{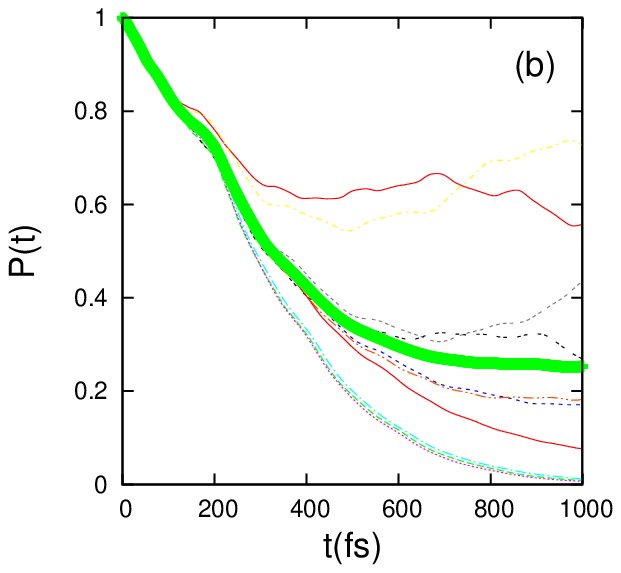}
\end{minipage}
\begin{minipage}{.42\linewidth}
\includegraphics[scale=.7]{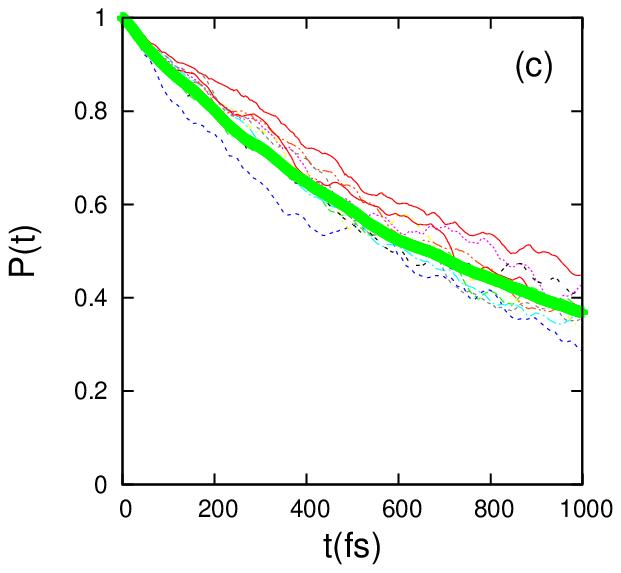}
\end{minipage}
\hspace{-2cm}
\begin{minipage}{.42\linewidth}
\includegraphics[scale=.7]{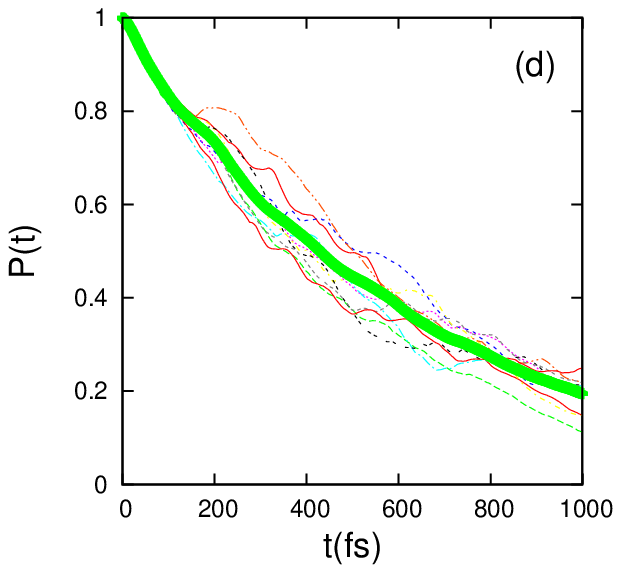}
\end{minipage}
\end{center}
\caption{\baselineskip5mm
Amide I mode relaxation probability $P(t)$ of NMA, as
  obtained from an instantaneous normal mode analysis with (right) and
  without (left) partial energy minimization. Shown are results from
  (upper panels) the inhomogeneous averaging approximation and (lower
  panel) dynamics averaging.  The VER times according to
  Eq.\ (\ref{eq:CGrate}) are (a) $\tau=0.54 \pm 0.23$ ps, (b)
  $\tau=0.43 \pm 0.15$ ps, (c) $\tau=0.93 \pm 0.15$ ps, and (d)
  $\tau=0.60 \pm 0.16$ ps.  }
\label{fig:NMAd1}
\end{figure}

We are now in a position to study the outcome of the various
theoretical treatments of VER introduced above.  To this end,
Fig.~\ref{fig:NMAd1} shows the VER probability $P(t)$ of the amide I
mode for {ten} representative trajectories (thin line) as well as the
mean population averaged {over 60 trajectories} (thick line). We again
compare results obtained from normal mode calculations with (right)
and without (left) partial geometry optimization of the subsystem NMA.
Furthermore, the dynamic treatment [Eq.\ (\ref{eq:VERexpo3})] as well
as its inhomogeneous limit [Eq.\ (\ref{eq:VERexpo2})] are
considered. In all cases, we find three stages of the time evolution
of $P(t)$. First, we observe a quadratic initial decay for $t
\lesssim$ 10 fs, which can be directly derived as short-time
expansion of Eq.\ (\ref{eq:VERexpo2}). This is followed by an
approximately linear decay for times up to $\approx$ 200
fs. Corresponding to a frequency uncertainty of $\Delta \omega
\approx$ 50 cm$^{-1}$, this is the time scale on which the normal mode
spectrum $\{ \omega_i \}$ of the ``total'' system
(NMA/(D$_2$O)$_{16}$) appears continuous, because $\Delta \omega
\gtrsim \omega_S -\omega_{\alpha} - \omega_{\beta}$. At longer times,
the discrete nature of the bath becomes apparent, leading to 
dispersion of the population probability for various MD trajectories.

As may be expected, the different time dependence of the normal mode
frequencies with and without the partial minimization also results in
different time evolutions of the VER probability. However, the
stronger fluctuations of the frequencies and couplings obtained from
the calculations without minimization (see Fig.\ \ref{fig:amideI}) do
not necessarily result in enhanced fluctuations of $P(t)$. More
important is whether the time-dependent fluctuations are taken into
account in the perturbative calculation. After $\gtrsim$ 100 fs, we
observe that the inhomogeneously approximated VER probabilities
largely disperse, while the dynamic treatment leads to comparatively
similar population decays for various MD trajectories. This is because
in the former case we use a constant resonance condition $\omega_S =
\omega_{\alpha} + \omega_{\beta}$, which may differ significantly for
individual trajectories. In the dynamic treatment, on the other hand,
the time integration in Eq.~(\ref{eq:VERexpo3}) smoothens the
time-dependent resonance condition $\omega_S(t) = \omega_{\alpha}(t) +
\omega_{\beta}(t)$.  To roughly quantify the VER dynamics, we
introduce an average time scale $\tau$ via\cite{FS07}
\begin{equation} \label{eq:CGrate}
\frac{1}{\tau} = - \frac{1}{t_2-t_1}\int_{t_1}^{t_2} \frac{d
  P(t)}{dt}\, dt = - \frac{ P(t_2) - P(t_1)}{t_2-t_1} ,
\end{equation}
with $t_1=0.0$ ps and $t_2$=0.5 ps. (Quite similar results are
obtained for $t_1=0.5$ ps and $t_2$=1.0 ps.) As noted in the caption
of Fig.~\ref{fig:NMAd1}, we find that the relaxation time $\tau$ as
well as its fluctuation may depend significantly on the theoretical
treatment of VER.

\begin{figure}[htbp]
\hfill
\begin{center}
\begin{minipage}{.42\linewidth}
\includegraphics[scale=.7]{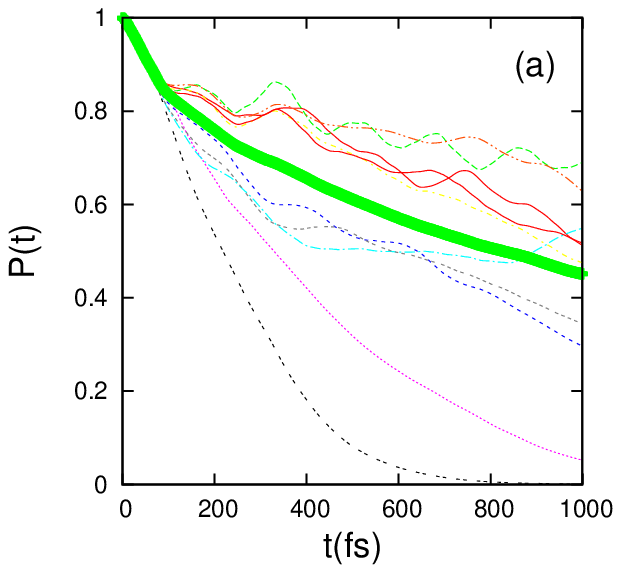}
\end{minipage}
\hspace{-2cm}
\begin{minipage}{.42\linewidth}
\includegraphics[scale=.7]{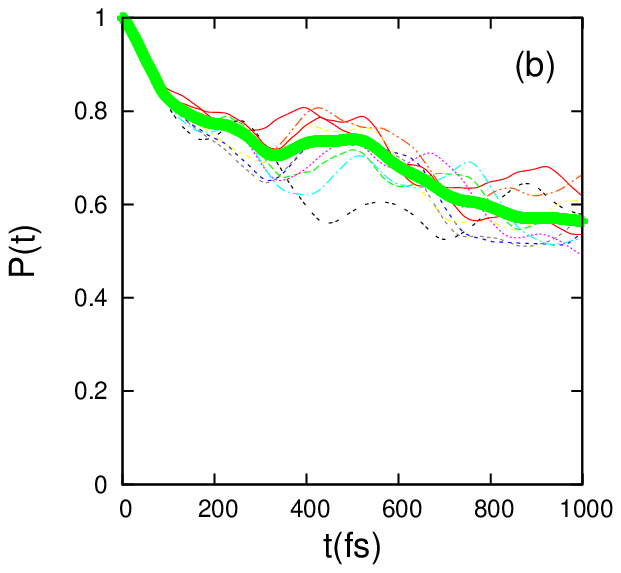}
\end{minipage}
\begin{minipage}{.42\linewidth}
\includegraphics[scale=.7]{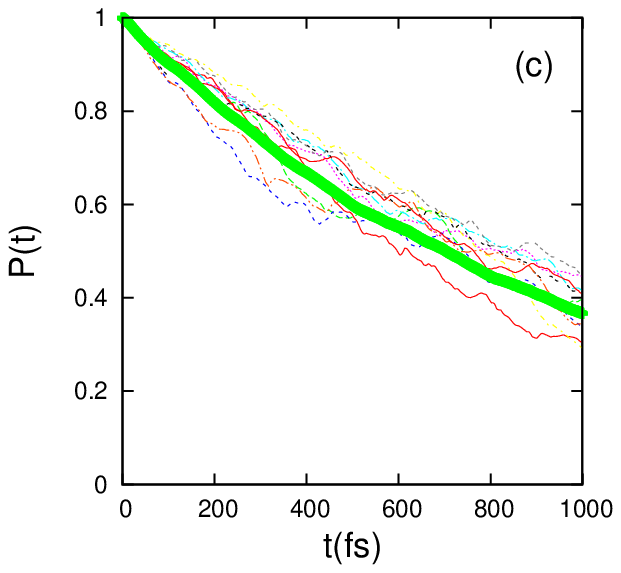}
\end{minipage}
\hspace{-2cm}
\begin{minipage}{.42\linewidth}
\includegraphics[scale=.7]{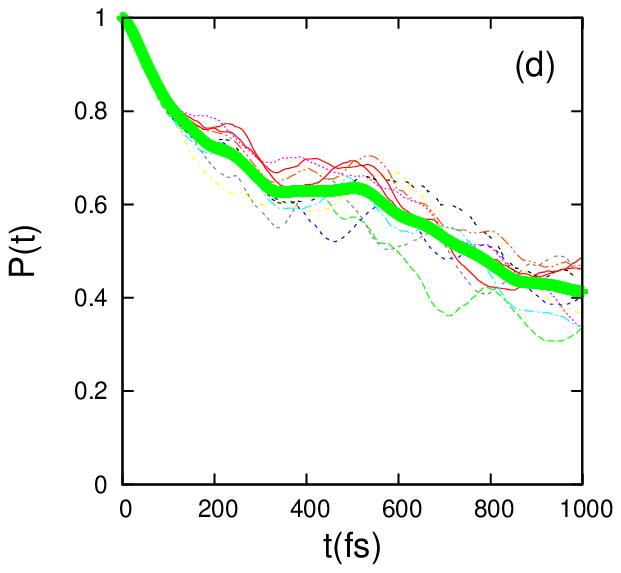}
\end{minipage}
\end{center}
\caption{\baselineskip5mm
Same as in Fig.\ \ref{fig:NMAd1}, but for fully deuterated
NMA.
 The VER times are 
(a) $\tau=0.89 \pm 0.69$ ps,
(b) $\tau=1.64 \pm 0.32$ ps,
(c) $\tau=0.96 \pm 0.28$ ps, and 
(d) $\tau=1.09 \pm 0.32$ ps.
}
\label{fig:NMAd7}
\end{figure}

It is instructive to compare the above results with calculations for
fully deuterated N-methylacetamide. As shown in Fig.\ \ref{fig:NMAd7},
deuteration significantly affects the relaxation. While there are
again clear differences between the various theoretical treatments,
the trends are not that clear as in the case of singly deuterated NMA
shown in Fig.\ \ref{fig:NMAd1}. Since deuteration may change normal
mode frequencies as well as vibrational couplings, this rises the
question on the relaxation pathway underlying the VER of NMA.

\squeezetable
\begin{table}[htbp]
\caption{\baselineskip5mm
Dominant energy flow pathways of singly deuterated NMA (upper
  panel) and fully deuterated NMA (lower panel). The vibrational
  energy relaxation from of the initially excited amide I mode
  ($\omega_S$) to the vibrational modes $\omega_\alpha$ and
  $\omega_\beta$ is characterized by their path-specific relaxation
  times $\tau_{S \alpha \beta}$ (in ps), time-averaged Fermi resonance
  parameter $\langle F_{S \alpha \beta} \rangle$, frequency mismatch
  $\langle \Delta \omega_{S \alpha \beta} \rangle$ (in cm$^{-1}$), and
  3rd order coupling element $\langle C_{S \alpha \beta} \rangle$ (in
  kcal/mol/\AA), respectively.}
\vspace{1cm}
\begin{tabular}{c|c|c|c|c|c|c|c|c|c}
\hline \hline
Mode $\alpha, \beta$ & 109,116 & 108,116 & 109,115 & 109,117 & 110,116 & 112,112 & 108,117 & 108,114 & 109,114 
\\ \hline           
$\tau_{S \alpha \beta}$ & 3.3 & 3.6 & 6.5 & 7.1 & 17.5 & 18.7 & 21.0 & 43.2 & 44.7
\\ \hline
$\langle F_{S \alpha \beta} \rangle$ & 0.209 & 0.176 & 0.109 & 0.043 & 0.033 & 0.046 & 0.055 & 0.030 & 0.028
\\ \hline
$\langle \Delta \omega_{S \alpha \beta} \rangle$ & 9.5 & 17.9 & 14.6 & 37.8 & 164.6 & 46.0 & 21.5 & 56.0 & 37.0
\\ \hline  
$\langle C_{S \alpha \beta} \rangle$ & 2.7 & 4.2 & 2.1 & 2.3 & 8.4 & 3.1 & 1.6 & 2.2 & 1.4
\\ \hline           
\end{tabular}
\vspace*{4mm}

\begin{tabular}{c|c|c|c|c|c|c|c|c|c}
\hline \hline
Mode $\alpha, \beta$ & 116,116 & 109,116 & 110,116 & 108,116 & 109,119 & 081,109 & 076,116 & 079,116 & 075,109 
\\ \hline           
$\tau_{S \alpha \beta}$ & 24.6 & 26.3 & 33.5 & 57.7 & 78.7 & 80.9 & 107.2 & 108.9 & 142.8
\\ \hline           
$\langle F_{S \alpha \beta} \rangle$ & 0.046 & 0.054 & 0.096 & 0.017 & 0.008 & 0.002 & 0.003 & 0.002 & 0.002
\\ \hline
$\langle \Delta \omega_{S \alpha \beta} \rangle$ & 252.3 & 151.2 & 60.1 & 172.7 & 94.6 & 937.0  & 565.4   & 546.9  & 974.7
\\ \hline  
$\langle C_{S \alpha \beta} \rangle$ & 18.9  & 10.1 & 7.7 & 3.6 & 0.93& 0.91 & 1.2 & 1.0 & 0.9
\\ \hline           
\end{tabular}
\label{table:FRP1}
\end{table}

Following definition (\ref{eq:CGrate}) of an average relaxation time
scale $\tau$, the individual energy flow pathways can be characterized
by introducing a path-specific VER decay times $\tau_{S \alpha
\beta}$. (Note that $1/{\tau} =\sum_{\alpha \beta} 1/{\tau_{S \alpha
\beta}}$.)  Adopting again the sample trajectory shown in Fig.\
\ref{fig:amideI}, Table \ref{table:FRP1} lists the dominant relaxation
pathways characterized by the shortest $\tau_{S \alpha
\beta}$. In the case of singly
deuterated NMA, two mode combinations [($\alpha, \beta)$ = (108,116)
and (109,116)] results in particularly fast relaxation. To further
analyze the relaxation mechanism, we consider the {time-averaged} 
Fermi resonance parameter defined as\cite{FYHS07,FYHSS08}
\begin{eqnarray} \label{eq:FRP}
\langle F_{S \alpha \beta} \rangle &=&
\left \langle
\sqrt{\frac{\hbar}{8 {\omega}_S \omega_{\alpha} \omega_{\beta}}}
\right \rangle
\frac{ \langle |C_{S \alpha \beta}| \rangle}
{\langle |\Delta \omega_{S \alpha \beta}| \rangle} ,
\end{eqnarray}
where $\langle \ldots \rangle$ denotes the time average over a single
trajectory. We see that in most (but not all) cases a large Fermi
resonance parameter $\langle F_{S \alpha \beta} \rangle$ also means a
short corresponding relaxation time $\tau_{S \alpha
\beta}$. Furthermore, it is interesting to note that it is the
resonance factor $\Delta \omega_{S \alpha \beta}$ rather than the
anharmonic coefficient $C_{S \alpha \beta}$ that
mostly determines $F_{S\alpha\beta}$ and thus the relaxation pathway.\cite{FYHS07,FYHSS08}

In the case of fully deuterated NMA, on the other hand, there are no
such strongly resonant modes. Instead we find that the subpicosecond
decay of the overall VER probability $P(t)$ is the result of numerous
decay channels with relatively long path-specific decay times. The
weak oscillations seen in Fig.\ \ref{fig:NMAd7} reflect the detuning
$\Delta \omega_{S \alpha \beta} \approx$ 200 cm$^{-1}$ of the two most
important paths.  Although quite similar in simulation and
experiment,\cite{HLH98,ZAH01,Tokmakoff06} interestingly, the mechanism
of amide I VER is different for singly and fully deuterated NMA.

\squeezetable
\begin{table}[h]
\caption{\baselineskip5mm
Characterization of the normal modes that mainly participate
  in the vibrational energy relaxation of singly deuterated NMA (upper
  panel) and fully deuterated NMA (lower panel). Shown are the
  vibrational frequency $\omega_{\alpha}$ and various projections on
  atomic coordinates, see Eq.\ (\ref{eq:Proj}).  Mode 141 is the amide
  I mode.}
\vspace{1cm}
\begin{tabular}{c|c|c|c|c|c|c|c|c|c|c|c|c}
\hline \hline
Mode \# &  107 & 108 & 109 & 110 & 111 & 112 & 113 & 116 & 117 & 139 & 140 & 141 
\\ \hline  
$\langle \omega_{\alpha} \rangle$ (cm$^{-1}$) & 454 & 571 & 590 & 752 & 765 & 864 & 983 & 1093 & 1129 & 1448 & 1570 & 1681
\\ \hline            
$P_{\rm CO}$ & 0.16 & 0.25 & 0.27 & 0.39 & 0.37 & 0.00  & 0.12 & 0.15 & 0.19 & 0.36 & 0.09 & 0.92
\\ \hline            
$P_{\rm ND}$ & 0.10 & 0.51 & 0.33 & 0.28 & 0.39 & 0.63 & 0.04 & 0.11 & 0.39 & 0.16 & 0.15 & 0.04
\\ \hline            
$P_{\rm CH_3(C)}$ & 0.28 & 0.18 & 0.31 & 0.27 & 0.16 & 0.00 & 0.83 & 0.32 & 0.18 & 0.22 & 0.01 & 0.04
\\ \hline            
$P_{\rm CH_3(N)}$ & 0.16 & 0.05 & 0.07 & 0.05 & 0.08 & 0.36 & 0.02 & 0.42 & 0.24 & 0.26 & 0.75 & 0.00
\\ \hline            
$P_{\rm Water}$ & 0.31 & 0.02 & 0.02 & 0.00 & 0.00 & 0.00 & 0.00 & 0.00 & 0.00 & 0.00 & 0.00 & 0.00
\\ \hline           
\end{tabular}
\vspace*{4mm}

\begin{tabular}{c|c|c|c|c|c|c|c|c|c|c|c|c}
\hline \hline
Mode \# & 107 & 108 & 109 & 110 & 111 & 112 & 113 & 116 & 117 & 139 & 140 & 141 
\\ \hline  
$\langle \omega_{\alpha} \rangle$ (cm$^{-1}$) & 433 & 549 & 560 & 651 & 686 & 787 & 798 & 963 & 1010 & 1293 & 1485 & 1675
\\ \hline            
$P_{\rm CO}$ & 0.08 & 0.28 & 0.29 & 0.10 & 0.17 & 0.03 & 0.19 & 0.49 & 0.07 & 0.00 & 0.46 & 0.94
\\ \hline            
$P_{\rm ND}$ & 0.04 & 0.21 & 0.40 & 0.34 & 0.20 & 0.36 & 0.23 & 0.13 & 0.12 & 0.00 & 0.39 & 0.04
\\ \hline            
$P_{\rm CH_3(C)}$ & 0.12 & 0.40 & 0.25 & 0.41 & 0.23 & 0.37 & 0.42 & 0.30 & 0.43 & 0.00 & 0.05 & 0.02
\\ \hline            
$P_{\rm CH_3(N)}$ & 0.08 & 0.10 & 0.03 & 0.14 & 0.40 & 0.23 & 0.16 & 0.07 & 0.38 & 0.00 & 0.11 & 0.00
\\ \hline            
$P_{\rm Water}$ & 0.69 & 0.02 & 0.03 & 0.01 & 0.00 & 0.00 & 0.00 & 0.00 & 0.00 & 1.00 & 0.00 & 0.00
\\ \hline           
\end{tabular}
\label{table:norm1}
\end{table}

To characterize the normal modes that mainly participate in the VER
process, we have calculated their projection on various atoms of NMA
and the surrounding water molecules. For example, the projection of
normal mode $\alpha$ onto the CO bond is defined by
\begin{equation} \label{eq:Proj}
P_{\rm CO}=\sum_{i \in {\rm CO \, bond}} U_{i \alpha}^2 ,
\end{equation}
where the sum goes over the $x,y,$ and $z$ coordinates of the
corresponding C and O atoms, respectively, and $\{U_{i \alpha}\}$
comprises the eigenvectors of the Hessian matrix [see Eq.~(\ref{eq:CSab})]. 
Listing these
projections, Table \ref{table:norm1} shows that the resonant modes are
mainly localized on NMA, rather than on the solvent water. 
We also notice that deuteration red-shifts some of the lower
frequencies, as expected, but not the amide I mode frequency.  This
appears to be the main reason of the little resonance of the
fully deuterated case. 

\begin{figure}[htbp]
\hfill
\begin{center}
\begin{minipage}{.32\linewidth}
\includegraphics[scale=0.7]{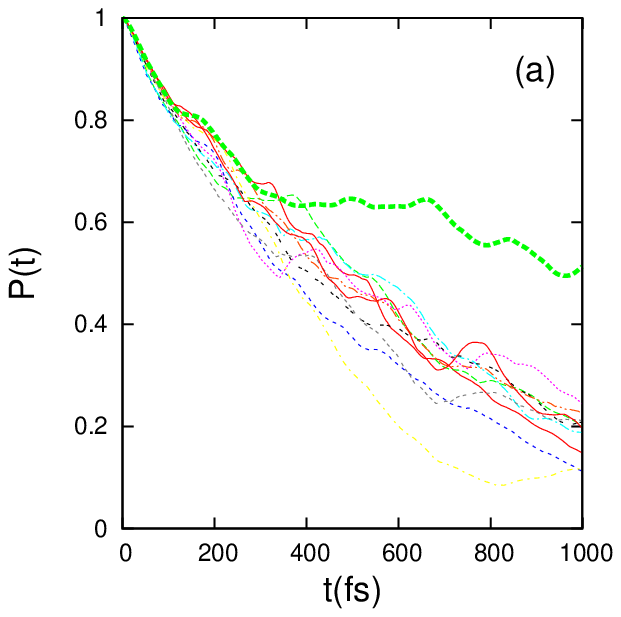}
\end{minipage}
\hspace{-1cm}
\begin{minipage}{.32\linewidth}
\includegraphics[scale=0.7]{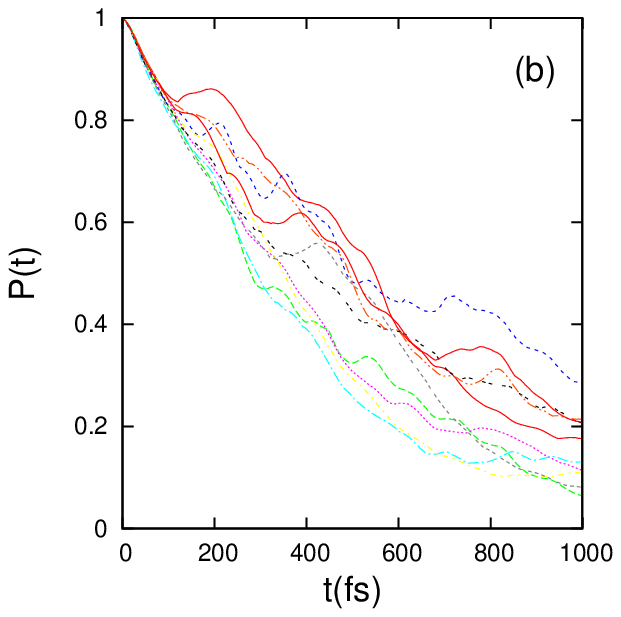}
\end{minipage}
\hspace{-1cm}
\begin{minipage}{.32\linewidth}
\includegraphics[scale=0.7]{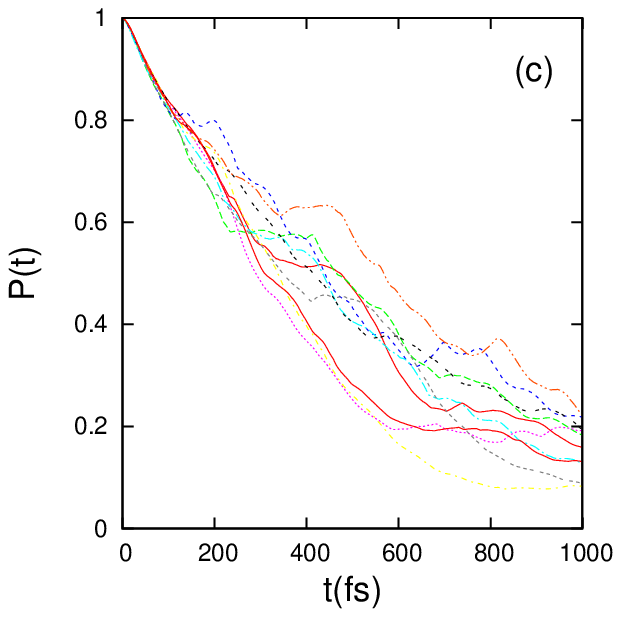}
\end{minipage}
\end{center}
\caption{\baselineskip5mm
Influence of the solvent on the amide I mode relaxation in
  NMA. Including 16 (left), 32 (middle), or all (right) water
  molecules in the calculations, the resulting overall vibrational
  energy relaxation appears to be quite similar. In the case of
  isolated NMA (thick line in left panel), however, the relaxation
  probability $P(t)$ follows the solution-phase decay only up to
  $\approx$ 300 fs. Note that in the absence of solvent molecules, all
  trajectories coincide.}
\label{fig:comp}
\end{figure}

To further study the influence of the solvent, Fig.~\ref{fig:comp}
compares the VER dynamics of NMA, when various numbers of D$_2$O
molecules are included in the normal mode and subsequent VER
calculations. In all cases, we show the VER probability of 10 single
trajectories for singly deuterated NMA, obtained from the dynamic VER
treatment and partial energy minimization. 
{Including 16 and 32 water molecules}, 
the resulting amide I
relaxation appears to be quite similar [Fig.~\ref{fig:comp} (a) and (b)]. 
{As another test, we furthermore 
took into account the remaining 535 water molecules as fixed charges [Fig.~\ref{fig:comp} (c).]}
Although individual relaxation
pathways may be different, the figure indicates that the first
solvation shell with 16 water molecules is sufficient to account for the first
step of VER of the amide I mode. Also shown in the left panel of
Fig.\ \ref{fig:comp} is the case of isolated NMA.  In the absence of
solvent water, we find that the excited state population
probability $P(t)$ follows the solution-phase decay only up to
$\approx$ 300 fs, and deviates to a larger value at longer times. This
is a consequence of the fact that while at short times a few
resonant pathways dominate, there are numerous possibilities of
``off-resonant pathways'' that may contribute at longer times.

\begin{figure}[htbp]
\hfill
\begin{center}
\begin{minipage}{.42\linewidth}
\includegraphics[scale=0.7]{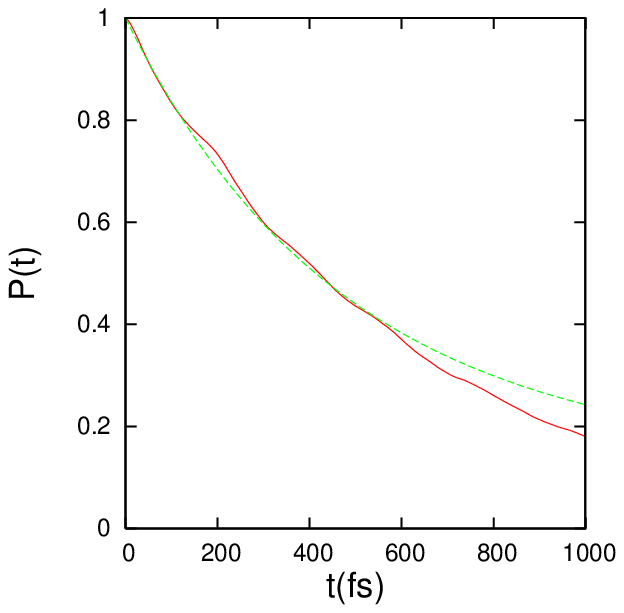}
\end{minipage}
\hspace{-2cm}
\begin{minipage}{.42\linewidth}
\includegraphics[scale=0.7]{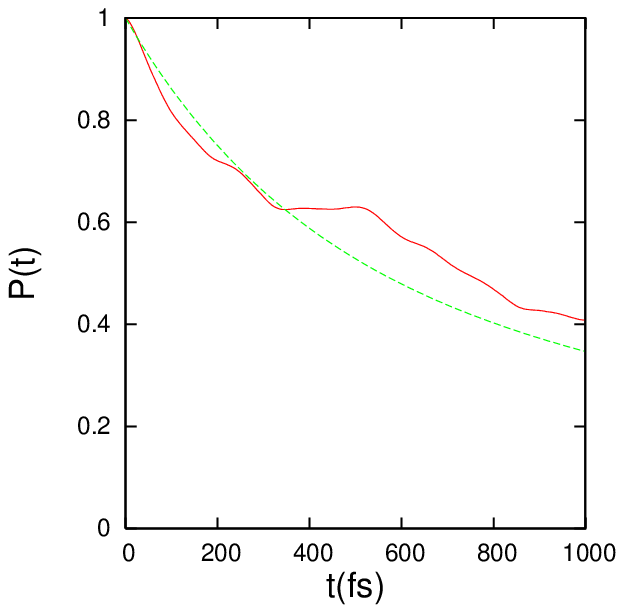}
\end{minipage}
\end{center}
\caption{\baselineskip5mm
Comparison of calculated (red lines) and experimental (green
lines) results for the VER of the amide I mode in NMA. Shown are data
for singly (left) and fully (right) deuterated NMA.}
\label{fig:exp}
\end{figure}

Finally we remark on the comparison of the above results with existing
experimental data. For both singly and fully deuterated NMA, biphasic
relaxation $P_{\rm exp}(t) = p e^{-t/\tau_1} + (1-p) e^{-t/\tau_2}$ of
the amide I mode has been reported.\cite{HLH98,ZAH01,Tokmakoff06}
The relaxation times are $\tau_1$ = 0.45 ps ($p=0.8$) and $\tau_2$ = 4 ps for
the singly deuterated NMA,\cite{ZAH01} and $\tau_1$ = 0.38 ps ($p=0.5$) and
$\tau_2$ = 2.1 ps for the fully deuterated NMA.\cite{Tokmakoff06}
Comparing $P_{\rm exp}(t)$ with our results obtained from the dynamic
VER treatment and partial energy minimization, Fig. \ref{fig:exp}
shows excellent agreement between theory and experiment -- maybe
better as it can be expected from a simple force field modeling of the
normal modes of the system.  Independent of force field uncertainties,
however, is our finding that the VER of the amide I mode in NMA
consists of two phases (see also Figs.~\ref{fig:NMAd1} and \ref{fig:NMAd7}). 
This is because
for $t\lesssim$ 200 fs the system's spectral density appears
continuous due to the frequency-time uncertainty relation, while at
longer times the discrete nature of the bath becomes apparent.
Although it is tempting to speculate that this behavior can explain
the experimentally obtained biphasic relaxation of NMA, it is at
present not clear if the approximations involved in our description
allow for this conclusion. (E.g., the validity of time-dependent
perturbation theory deteriorates at longer times.) Nevertheless, the
observation that amide I relaxation in larger peptides occurs in a
monoexponential manner\cite{HLH98} is consistent with the fact that
the above described effect is expected to vanish for larger peptides
with higher spectral density.

\clearpage
%
%
\section{Concluding Remarks}
\label{sec:summary}

We have outlined a computational approach to describe the energy
relaxation of a high-frequency vibrational mode in a fluctuating
heterogeneous environment. Extending the previous work,\cite{FZS06} we
have employed second time-dependent perturbation theory, which
includes the fluctuations of the parameters in the Hamiltonian within
the vibrationally adiabatic approximation. This means that the
time-dependent vibrational frequencies along an MD trajectory are
obtained via a partial geometry optimization of the peptide with fixed
solvent water and a subsequent normal mode calculation. Although it
requires more computational effort, the partial geometry optimization
is necessary, because its omission results in unrealistically high
fluctuations of the vibrational frequencies and other quantities 
(see Fig. \ref{fig:amideI}).

Adopting the amide I VER of NMA in heavy water as a test problem, we
have shown that the inclusion of dynamic fluctuations may
significantly change the time evolution of the VER probability $P(t)$
(see Fig.\ \ref{fig:NMAd1}). After $\gtrsim$ 200 fs, we observe that
the inhomogeneously approximated VER probabilities largely disperse,
while the dynamic treatment leads to comparatively similar population
decays for various MD trajectories. This is because in the former case
we use a constant resonance condition $\omega_S = \omega_{\alpha} +
\omega_{\beta}$, while the time integration in the dynamic treatment 
averages over the time-dependent resonance condition.

To characterize the dominant energy flow pathways of the amide I VER
of NMA, we have introduced path-specific VER decay times and
considered the Fermi resonance parameter of the relaxation process. In
the case of singly deuterated NMA, mainly two combinations of bath
modes were found to achieve the vibrational relaxation. In the case of
fully deuterated NMA, on the other hand, we have not observed such
strongly resonant modes. Instead we found that the subpicosecond decay
of the overall VER probability $P(t)$ is the result of numerous decay
channels with relatively long path-specific decay times. In both
cases, we observed that the VER of the amide I mode in NMA consists of
two phases (see Fig.\ \ref{fig:NMAd1}). This is because for
$t\lesssim$ 200 fs the system's spectral density appears continuous
due to the frequency-time uncertainty relation, while at longer times
the discrete nature of the bath becomes apparent. Considering our
excellent agreement between theory and experiment (see Fig.\
\ref{fig:exp}), it may be speculated if this behavior can explain the
experimentally obtained biphasic relaxation of NMA.

Two directions for further researches are apparent. On a short time
scale (here $\lesssim$ 200 fs), our results suggest that a single
normal mode calculation should be sufficient to describe VER in
peptides. The restriction to ultrashort time scales therefore allows
us to employ high-quality {\em ab initio} methods to calculate the
vibrational structure of even large molecules.\cite{FYHS07,FYHSS08} 
On a longer time
scale (here $\gtrsim$ 500 fs), on the other hand, the validity of
time-dependent perturbation theory is expected to deteriorate. As
a straightforward extension, the vibrational dynamics may be described
by stochastic Schr\"odinger equations \cite{GNKS07} or 
stochastic Liouville equations.\cite{Hanggi,Tanimura06}

\acknowledgments
We thank Phuong H. Nguyen, Sang-Min Park, Tobias Brandes, and John E. Straub 
for discussions and helpful comments. 
HF is grateful to the Alexander von
Humboldt Foundation for their generous support. This work has been
supported by the Frankfurt Center for Scientific Computing and the
Fonds der Chemischen Industrie.

\newpage
%
%


\begin{thebibliography}{99}

\bibitem{Agarwal}
P.K. Agarwal, S.R. Billeter, P.R. Rajagopalan, S.J. Benkovic, and 
S. Hammes-Schiffer, 
Proc. Natl. Acad. Sci. USA {\bf 301}, 2794 (2002); 
P. Agarwal, J. Am. Chem. Soc. {\bf 127}, 15248 (2005); 
A. Jim\'enez, P. Clap\'es, and R. Crehuet,
J. Mol. Model (2008), doi:10.1007/s00894-008-0283-2.


\bibitem{Steinfeldbook}
J.I. Steinfeld, J.S. Francisco, and W.L. Hase,
{\it Chemical Kinetics and Dynamics},
Prentice-Hall (1989).


\bibitem{LW97}
D.M. Leitner and P.G. Wolynes,
Chem. Phys. Lett. {\bf 280}, 411 (1997). 

\bibitem{GW04}
M.~Gruebele and P.G.~Wolynes,
Acc.~Chem.~Res.~{\bf 37}, 261 (2004);
M. Gruebele,
J. Phys.: Condens. Matter {\bf 16}, R1057 (2004). 



\bibitem{MillerReview}
R.J.D. Miller,
Ann. Rev. Phys. Chem. {\bf 42}, 581 (1991);
R.J. Dwayne Miller, 
Can. J. Chem. {\bf 80}, 1 (2002); 
A.M. Nagy, V.I. Prokhorenko, R.J.D. Miller, 
Curr. Opin. Struct. Biol. {\bf 16}, 654 (2006).

\bibitem{MK97}
Y. Mizutani and T. Kitagawa,
Science {\bf 278}, 443
(1997);
Y. Mizutani and T. Kitagawa,
Chem. Record {\bf 1}, 258 (2001). 

\bibitem{HLH98}
P.~Hamm, M.H.~Lim, R.M.~Hochstrasser,
J.~Phys.~Chem.~B {\bf 102}, 6123 (1998). 

\bibitem{ZAH01}
M.T.~Zanni, M.C.~Asplund, R.M.~Hochstrasser,
J.~Chem.~Phys.~{\bf 114}, 4579 (2001). 

\bibitem{Tokmakoff06}
L.P.~DeFlores, Z.~Ganim, S.F.~Ackley, H.S.~Chung, A.~Tokmakoff,
J.~Phys.~Chem.~B {\bf 110}, 18973 (2006).

\bibitem{FayerReview}
M.D. Fayer,
Ann. Rev. Phys. Chem. {\bf 52}, 315 
(2001).


\bibitem{DRID00}
J.C. De\`ak, S.T. Rhea, L.K. Iwaki, and D.D. Dlott,
J. Phys. Chem. A {\bf 104}, 4866
(2000).


\bibitem{NNT05}
S. Nishida, T. Nada and M. Terazima,
Biophys. J. {\bf 89}, 2004 
(2005).


\bibitem{Hammreview}
S. Woutersen and P. Hamm,
J. Phys.: Condens. Matter {\bf 14}, R1035 (2002);
P. Hamm, J. Helbing, and J. Bredenbeck,
Annu. Rev. Phys. Chem. {\bf 59}, 291 (2008).

\bibitem{Oxtoby}
D.W.~Oxtoby, 
Adv.~Chem.~Phys.~{\bf 40}, 1 (1979);
Adv. Chem. Phys.~{\bf 47}, 487 (1981);
Ann. Rev. Phys. Chem. {\bf 32}, 77 (1981).


\bibitem{KTF94}
V.M. Kenkre, A. Tokmakoff, and M.D. Fayer,
J. Chem. Phys. {\bf 101}, 10618 (1994).

\bibitem{Hynes}
R.~Rey, K.B.~Moller, and J.T.~Hynes,
Chem.~Rev.~{\bf 104}, 1915 (2004);
K.B.~Moller, R.~Rey, and J.T.~Hynes,
J.~Phys.~Chem.~A {\bf 108}, 1275 (2004).

\bibitem{Skinner}
C.P.~Lawrence and J.L.~Skinner,
J.~Chem.~Phys.~{\bf 117}, 5827 (2002);
ibid.~{\bf 117}, 8847 (2002);
ibid.~{\bf 118}, 264 (2003); 
A.~Piryatinski, C.P.~Lawrence, and J.L.~Skinner, 
ibid.~{\bf 118}, 9664 (2003);
ibid.~{\bf 118}, 9672 (2003);
C.P.~Lawrence and J.L.~Skinner,
ibid.~{\bf 119}, 1623 (2003);
ibid.~{\bf 119}, 3840 (2003).

\bibitem{Sibert}
E.L. Sibert and R. Rey,
J. Chem. Phys. {\bf 116}, 237 (2002); 
T.S. Gulmen and E.L. Sibert,
J. Phys. Chem. A {\bf 109}, 5777 (2005); 
S.G. Ramesh and E.L. Sibert,
J. Chem. Phys. {\bf 125} 244512 (2006);
S.G. Ramesh and E.L. Sibert,
ibid.~{\bf 125} 244513 (2006).

\bibitem{Marx04}
R. Ramirez, T. Lopez-Ciudad, P. Kumar, and D. Marx,
J.~Chem.~Phys.~{\bf 121}, 3973 (2004).



\bibitem{Straub}
D.E.~Sagnella and J.E.~Straub,
Biophys.~J.~{\bf 77}, 70 (1999); 
L.~Bu and J.E.~Straub,
ibid.~{\bf 85}, 1429 (2003).


\bibitem{WWH92}
R.M. Whitnell, K.R. Wilson, and J.T. Hynes
J. Chem. Phys. {\bf 96}, 5354 (1992).

\bibitem{Henry}
E.R. Henry, W.A. Eaton, R.M. Hochstrasser,
Proc. Natl. Acad. Sci. USA 
{\bf 83}, 8982 (1986).


\bibitem{Straub2}
D. E. Sagnella and J. E. Straub, 
J. Phys. Chem. B {\bf 105}, 7057 (2001);
L. Bu and J. E. Straub, 
ibid.~{\bf 107}, 10634 (2003);
L. Bu and J. E. Straub,
ibid.~{\bf 107}, 12339 (2003);
Y. Zhang, H. Fujisaki, and J. E. Straub, 
ibid.~{\bf 111}, 3243 (2007).


\bibitem{NS03}
P.H.~Nguyen and G.~Stock,
J.~Chem.~Phys.~{\bf 119}, 11350 (2003);
P. H. Nguyen and G. Stock, 
Chem. Phys. {\bf 323}, 36 (2006);
P. H. Nguyen, R. D. Gorbunov, and G. Stock, 
Biophys. J. {\bf 91}, 1224 (2006);
A. Moretto, M. Crisma, C. Toniolo,  P. H. Nguyen, G. Stock, and 
P. Hamm, 
Proc. Natl. Acad. Sci. USA {\bf 104}, 12749 (2007). 

%
\bibitem{Moritsugu2000}
K. Moritsugu, O. Miyashita and A. Kidera,
Phys. Rev. Lett. {\bf 85}, 3970, (2000);
J. Phys. Chem. B, {\bf 107}, 3309 (2003).


\bibitem{Nagaoka}
I. Okazaki, Y. Hara, and M. Nagaoka,
Chem. Phys. Lett. {\bf 337}, 151 (2001); 
M. Takayanagi, H. Okumura, and M.Nagaoka,
J. Phys. Chem. B, {\bf 111}, 864 (2007).

\bibitem{Okazaki}
S. Okazaki,
Adv. Chem. Phys. {\bf 118}, 191 (2001);
M.~Shiga and S.~Okazaki,
J.~Chem.~Phys.~{\bf 109}, 3542 (1998);
ibid.~{\bf 111}, 5390 (1999);
T.~Mikami, M.~Shiga and S.~Okazaki,
ibid.~{\bf 115}, 9797 (2001);
T.~Terashima, M.~Shiga, and S.~Okazaki,
ibid.~{\bf 114}, 5663 (2001); 
Mikami, T.; Okazaki, S. 
ibid.~{\bf 119}, 4790 (2003);
T.~Mikami and S.~Okazaki,
ibid.~{\bf 121}, 10052 (2004);
M. Sato and S. Okazaki,
ibid.~{\bf 123}, 124508 (2005);
M. Sato and S. Okazaki  
ibid.~{\bf 123}, 124509 (2005).




\bibitem{Geva}
Q. Shi and E. Geva,
J. Chem. Phys. {\bf 118}, 7562 (2003); 
Q. Shi and E. Geva,
J. Phys. Chem. A {\bf 107}, 9059 (2003);
Q. Shi and E. Geva,
ibid.~{\bf 107}, 9070 (2003);
B.J. Ka, Q. Shi, and E. Geva,
ibid.~{\bf 109}, 5527 (2005);
B.J. Ka and E. Geva,
ibid.~{\bf 110}, 13131 (2006);
I. Navrotskaya and E. Geva,
ibid.~{\bf 111}, 460 (2007);
I. Navrotskaya and E. Geva,
J. Chem. Phys. {\bf 127}, 054504 (2007).


\bibitem{FBS05}
H.~Fujisaki, L.~Bu, and J.E.~Straub,
Adv.~Chem.~Phys.~{\bf 130}B, 
179 (2005); 
H.~Fujisaki and J.E.~Straub,
Proc.~Natl.~Acad.~Sci.~USA {\bf 102}, 
6726 (2005).


\bibitem{Leitner05}
D.M.~Leitner,
Adv.~Chem.~Phys.~{\bf 130}B, 
205 (2005);
D.M. Leitner, 
Phys. Rev. Lett. {\bf 87}, 188102 (2001);
X. Yu and D.M. Leitner,
J. Chem. Phys. {\bf 119}, 12673 (2003); 
X. Yu and D.M. Leitner,
J. Phys. Chem. B {\bf 107}, 1689 (2003);
D.M.~Leitner, M.~Havenith, and M.~Gruebele,
Int.~Rev.~Phys.~Chem.~{\bf 25}, 553 (2006).




\bibitem{DJBK07}
A.G. Dijkstra, T. la Cour Jansen, R. Bloem, and J. Knoester, 
J. Chem. Phys. {\bf 127}, 194505 (2007). 




\bibitem{FZS06}
H.~Fujisaki, Y.~Zhang, and J.E.~Straub,
J.~Chem.~Phys.~{\bf 124}, 144910 (2006).


\bibitem{FS07}
H.~Fujisaki and J.E.~Straub,
J. Phys. Chem. B {\bf 111}, 12017 (2007). 




\bibitem{FYHS07}
H.~Fujisaki, K.~Yagi, K.~Hirao, and J.E.~Straub,
Chem. Phys. Lett. {\bf 443}, 6 (2007). 



\bibitem{FYHSS08}
H.~Fujisaki, K.~Yagi, K.~Hirao, J.E.~Straub, and G.~Stock,
submitted.



\bibitem{Weiss}
U. Weiss,
{\it Quantum Dissipative Systems},
2nd ed. World Scientific (1999);
C.W. Gardiner and P. Zoller,
{\it Quantum Noise}, 3rd ed. Springer (2004);
D.F. Walls and G.J. Milburn,
{\it Quantum Optics}, 2nd ed. Springer (2008).


\bibitem{Breuer}
H.-P. Breuer and F. Petruccione, 
{\it The Theory of Open Quantum Systems},
Oxford University, New York (2002).

\bibitem{Nitzan}
A. Nitzan,
{\it Chemical Dynamics in Condesed Phase: 
Relaxation, Transfer, and Reactions in Condensed Molecular Systems}, 
Oxford (2006). 




\bibitem{Bakker04}
H. J. Bakker,
J. Chem. Phys. {\bf 121}, 10088 (2004).


\bibitem{Hanggi}
I. Goychuk and P. H\"anggi,
Adv. Phys. {\bf 54}, 525 (2005).


\bibitem{IT06} 
A. Ishizaki and Y. Tanimura 
J. Chem. Phys. {\bf 125}, 084501 (2006);
A. Ishizaki and Y. Tanimura 
ibid.~{\bf 123}, 014503 (2005).   


\bibitem{positivity}
T. Yu, L. Diosi, N. Gisin, and W.T. Strunz,
Phys. Lett. A {\bf 265}, 331 (2000);
G. Schaller and T. Brandes, 
e-print arXiv:0804.2374.






\bibitem{OT90}
I. Ohmine and H. Tanaka,
J. Chem. Phys. {\bf 93}, 8138 (1990);
I. Ohmine and H. Tanaka,
Chem. Rev. {\bf 93}, 2545 (1993). 

\bibitem{CHARMM}
B.R. Brooks, R.E. Bruccoleri, B.D. Olafson, D.J. States, S. Swaminathan, 
and M. Karplus,
J. Comp. Chem. {\bf 4}, 187 
(1983); 
A.D. MacKerell, Jr., B. Brooks, C.L. Brooks, III, L. Nilsson, B. Roux, Y. Won, and M. Karplus,
{\it The Encyclopedia of Computational Chemistry 1}. Editors, P.v.R. Schleyer et al. 
Chichester: John Wiley \& Sons. 271 
(1998). 



\bibitem{CHARMMFF}
A.D. MacKerell Jr., D. Bashford, M. Bellott, R.L. Dunbrack, J.D. Evanseck, M.J. Field,
S. Fischer, J. Gao, H. Guo, S. Ha, D. Joseph-McCarthy, L. Kuchnir, K. Kuczera, F.T.K. Lau,
C. Mattos, S. Michnick, T. Ngo, D.T. Nguyen, B. Prodhom, W.E. Reiher, B. Roux, M. Schlenkrich, 
J.C. Smith, R. Stote, J.E. Straub, M. Watanabe, J. Wiorkiewicz-Kuczera, D. Yin, and M. Karplus,
J. Phys. Chem. B {\bf 102}, 3586 (1998).

\bibitem{TIP3P}
W.L. Jorgensen, J. Chandrasekhar, J. Madura, R.W. Impey, and M.L. Klein,
J. Chem. Phys. {\bf 79}, 926 (1983). 

\bibitem{SHAKE}
J.-P. Ryckaert, G. Ciccotti, and H.J.C. Berendsen,
J. Comp. Phys. {\bf 23}, 327 (1977). 










\bibitem{GNKS07}
R.D. Gorbunov, P.H. Nguyen, M. Kobus, and G. Stock,
J. Chem. Phys. {\bf 126}, 054509 (2007);
M. Kobus, R. D. Gorbunov, P. H. Nguyen, and G. Stock,
Chem. Phys. {\bf 347}, 208 (2008). 











\bibitem{Tanimura06}
Y. Tanimura, 
J. Phys. Soc. Jpn. {\bf 75}, 082001  (2006).   





















\end{thebibliography}
\end{document}